\documentclass{aa} 
\usepackage{graphics}
 \thesaurus{06 (03.09.7; 08.03.4; 08.09.1; 08.09.2; 08.23.3)}
\title{Adaptive Optics imaging of P Cyg in H$_{\alpha}$ \thanks{Based
on data collected at the Observatoire de Haute-Provence}}
\author{Chesneau O.\inst{1} 
\and Roche M.\inst{2} 
\and Boccaletti A.\inst{3}
\and Abe L. \inst{1}
\and Moutou C.\inst{4}
\and Charbonnier F.\inst{5}
\and Aime C.\inst{2}
\and Lant\'eri H.\inst{2}
\and Vakili F.\inst{1}}
 \offprints{Chesneau O.} 
\mail{chesneau@astro.umontreal.ca}
 \institute{ Observatoire de
la C{\^o}te d'Azur, D{\'e}partement Fresnel, GI2T, F-06460 Saint
Vallier de Thiey
\and UMR 6525 Astrophysique, Universit{\'e} de Nice Sophia Antipolis Parc
Valrose, F-06108 Nice cedex 2
\and Coll{\`e}ge
de France, 11 Pl. M. Berthelot F-75321 Paris
\and European Southern Observatory, Alonso de Cordoba 3107, Santiago, Chile
\and Office National d'Etudes et de Recherches A{\'e}rospatiales, D\'epartement d'Optique Th\'eorique et
Appliqu\'ee Imagerie Haute R\'esolution - Optique Adaptative, 29 Av de
la Division Leclerc, F-92320 Chatillon}
 \date{Received January 10; accepted March 20, 2000} 
\authorrunning{Chesneau O. et al.}
 \begin{document}
\maketitle
\begin{abstract} 
 We obtained H$_{\alpha}$ diffraction limited data of the LBV star P Cyg
using the ONERA Adaptive Optics (AO) facility BOA at the OHP 1.52m telescope on October 1997.
Taking P Cyg and the reference star 59 Cyg AO long exposures we find that P Cyg clearly exhibits a
large and diffuse intensity distribution compared to the 59 Cyg's
point-like source. A deconvolution of P Cyg using 59 Cyg as the
Point Spread Function was performed by means of the Richardson-Lucy
algorithm. P Cyg clearly appears as an unresolved star surrounded by
a clumped envelope. The reconstructed image of P Cyg is compared to similar spatial resolution maps obtained from radio aperture synthesis imaging. We put independent constraints on the physics of P Cyg which agree well with radio results. 
We discuss future possibilities to constrain the wind
structure of P Cyg by using multi-resolution imaging, coronagraphy and long baseline interferometry
to trace back its evolutionary status.
\keywords{star: P Cyg, techniques: adaptive optics, deconvolution}
 \end{abstract}
 \section{Introduction}
Among the galactic Luminous Blue Variables (LBV), the supergiant P Cyg
(HD 193237, B1Ia+) is both a historical prototype since its famous
XVIIth century
eruptions, and an intriguing paradigm in many aspects. In their
exhaustive study of P Cyg's fundamental parameters, mass loss physics
and evolution, Nota et al. (\cite{nota95}) have held this star being
the unique case of the so-called "Peculiar Nebulae" class LBVs.
Indeed the 22 arcseconds (arcsec) faint and spherical nebulae around P Cyg were only recently
discovered by Barlow et al. (\cite{barlow94}) who estimate its mass as
0.01 solar mass. This value significantly differs from those of more
regular LBVs, ranging from 1$M_{\odot}$ to 4.2$M_{\odot}$
(e.g. for AG Car) which also present more asymmetric nebula.\\
\begin{figure}[t]
\centerline{\resizebox{7cm}{!}{\includegraphics{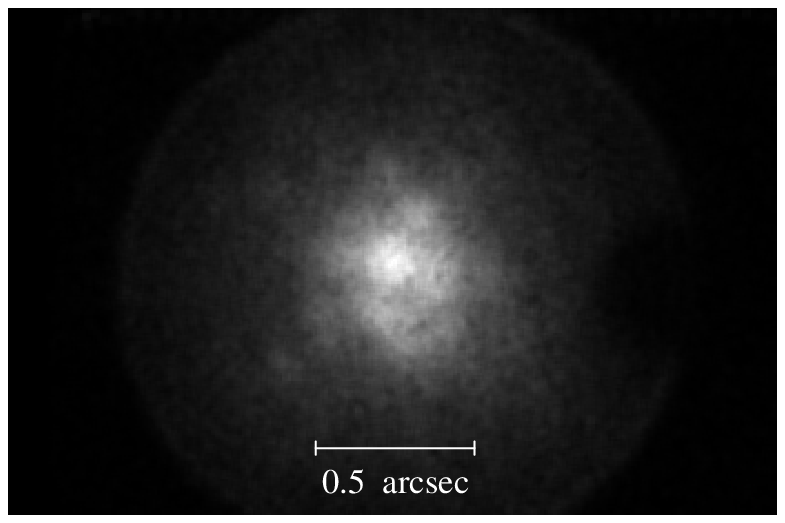}}}
\centerline{ \resizebox{7cm}{!}{\includegraphics{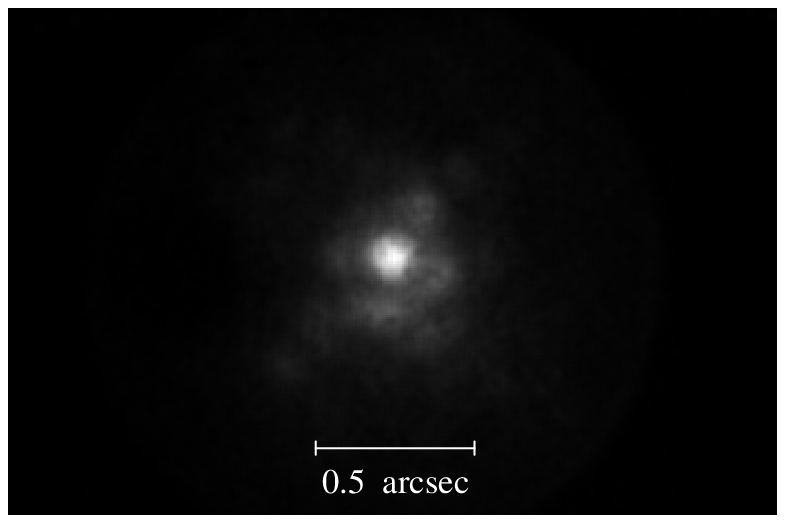}}}
\vspace{0.2cm} \caption[]{H$_{\alpha}$ long exposures of P Cyg (top)
  and the reference star 59 Cyg (bottom) recorded with the CP20 camera
  at OHP during October 4th 1997 using the AO system BOA from ONERA through a  filter ($\Delta\lambda$=10nm). The images are smoothed
using a moving average window of 4$\times$4 pixels. Note that the
theoretical angular resolution of the 1.52m telescope at 6563$\AA$  
(1.22$\lambda$/D=0.110 arcsec) is closely reached (broken first Airy ring is visible on 59 Cyg) whilst for P Cyg a more complex and resolved structure can be suspected.} \label{figun} \end{figure} 
Evolutive tracks 
suggest an initial mass of 48$\pm$6
$M_{\odot}$, and it's present mass is estimated to be at the most 40
$M_{\odot}$, but lower masses ($\sim$ 30 $M_{\odot}$) are also reported in literature
(\cite{lamers83}, \cite{lamers85}, and \cite{turner99}). However, the
fine spatial structure of this large amount of excreted matter remains
to be detailed.
P Cyg's relative proximity ($\sim$ 1.8 kpc, \cite{lamers83}) represents an 
opportunity
to observe its radiatively driven mass loss from the starting point out to 
the
interstellar medium. Indeed, at 1.8 kpc, the central star radius 
($76R_{\odot}$)
corresponds to a tiny angle of 0.2 milliarcsecond (mas) but the H$_{\alpha}$
emitting
region extends over several tens of arcsec, and radio emission 
seems
to
attain at even larger scales (\cite{meaburn99}).
The optical and radio observations depict so far an essentially clumpy
distribution of matter both at large (\cite{taylor91}, \cite{nota95}) and
small scales (\cite{skinner98}, \cite{vakili97}), 
with temporally variable emission (\cite{skinner97}). These  imaging observations 
remain
sparse and can loosely constrain the spatial and/or temporal evolution
of the clumps in the nebula.
Moreover, in the optical wavelengths, the star to the envelope brightness 
ratio
remains an obstacle for studying the central star
immediate
environment.
In this paper we report an attempt to observe the H$_{\alpha}$ circumstellar
environment
of P Cyg during an AO run at OHP observatory on October 1997 using
short exposures collected with a photon-counting camera.\\
\begin{figure}[t]
\centerline{\resizebox{7.5cm}{!}{\includegraphics{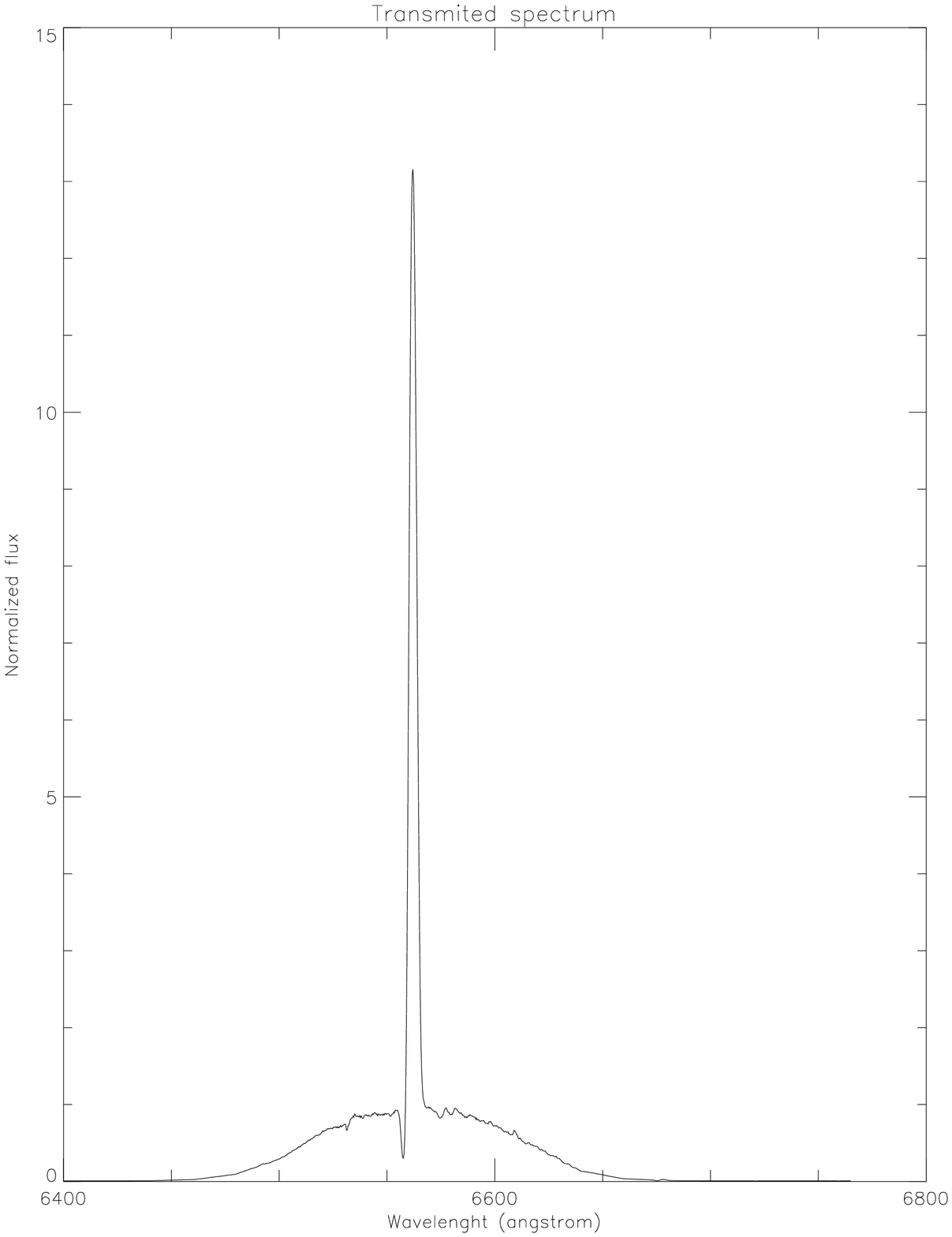}}}
\vspace{0.2cm} \caption[]{Synthetic representation of the $H_{\alpha}$
  region of P Cyg multiplied by the filter transmission used for the
  adaptive optics imaging of P Cyg. The same filter was used for 59
  Cyg. About 45.6\% of the transmitted light originates from the line continuum.} \label{figdeux} \end{figure} 
The paper is organized as follows. 
In section 2, we describe the AO+imaging instrumentation used for this
study as
well as data reduction and calibration procedures. Section 3 describes the 
image
reconstruction in which particular efforts have been made to test the 
validity
of
the PSF. The last section makes a critical discussion of our founding, and
attention
is given on the perspectives opened by the emergencies of new high angular
resolution
techniques. A need arise for a coordinated campaign for a global approach of P
Cyg,
and LBVs environment.

\section{Journal of observations} 
\subsection{Instrumentation} 
 P Cyg was observed, in the context of a dark-speckle run (\cite{boccaletti98}). The AO bench of the
ONERA (Office National d'Etudes et de Recherches A{\'e}rospatiales) was installed at the
Coud{\'e} focus of the 1.52m telescope. The 90 actuators of BOA  and
its 1 kHz closed loop bandwidth enable a compensation of
atmospheric turbulence in visible light (\cite{conan98}) and provide typical Strehl ratio of $10\%$ to
$30\%$ depending on seeing conditions.\\ 
The restored wavefront feeds a dark-speckle optics designed to detect faint companions around bright objects. The beam is then focused onto the
detector with an $f/976$ aperture, giving a fine pixel sampling of 144
$pixels/arcsec$ or $0.007arcsec/pixel$ (this oversampling of the images was
imposed for the dark-speckle experiment). The
detector is a cooled CP20 photon-counting camera (\cite{abe98}) allowing single photon detection
with a very low dark count ($<5.10^{-3}photon/pix/s$). The quantum efficiency of CP20 is less than
$10\%$ at $700nm$.\\ Near diffraction-limited images of P Cyg ($m_v=4.81$, $S_p=B1Ia+$) and 59 Cyg
($m_v=4.74$, $S_p=B1.5V$) have been obtained (Fig. \ref{figun}) using a broadband $H_{\alpha}$ filter
($\lambda_0=6563$\AA, $\Delta\lambda=100$\AA) (Fig. \ref{figdeux}).

\subsection{Data analysis} 

\begin{figure}[t]
\centerline{\resizebox{5.2cm}{!}{\includegraphics{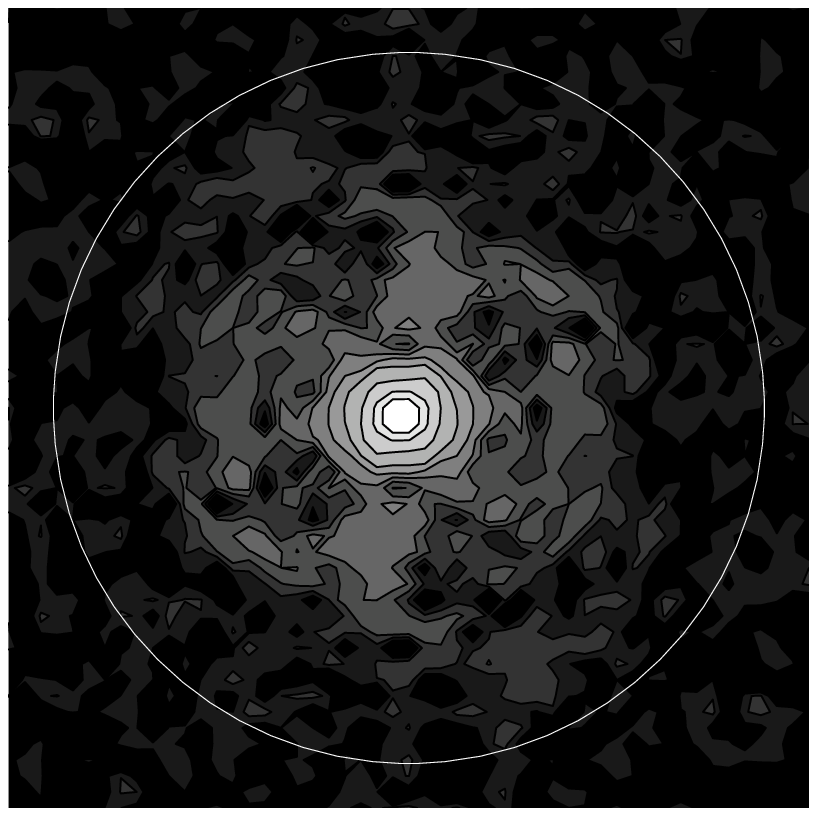}}}
\centerline{\resizebox{5.2cm}{!}{\includegraphics{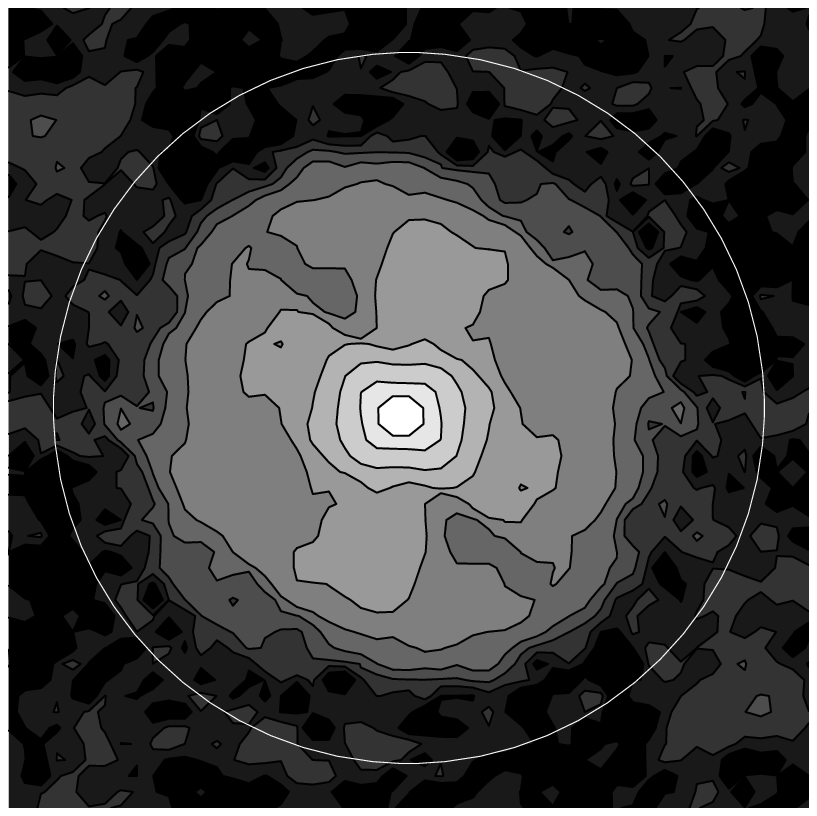}}}
\centerline{\resizebox{5.2cm}{!}{\includegraphics{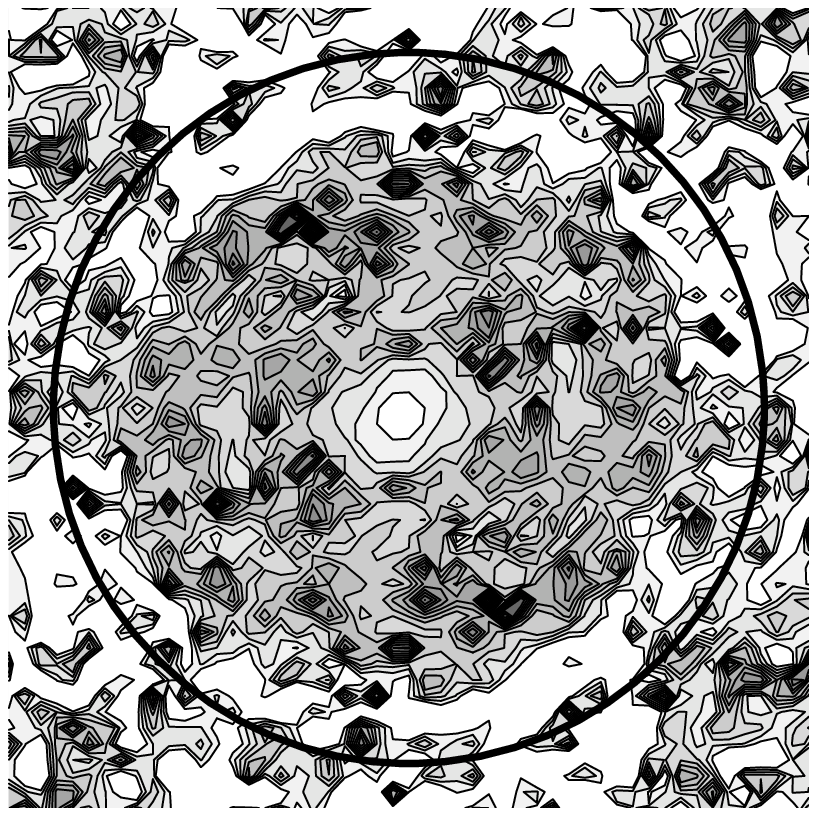}}}
\vspace{0.2cm} \caption[]{Representation in contour plots of the
  modulus of the Fourier transform of P Cyg (top), 59 Cyg (middle) and
  their ratio, the visibility curve of P Cyg (bottom). The zero
 frequency is at the center of the images. The representations are in
 logarithmic scale. The circle gives the theoretical cut-off frequency of the telescope.} \label{mod} \end{figure}

The data consist in two sequences of 20 ms short exposures, for respectively P Cyg, and the reference star
59 Cyg of similar magnitude and spectral type. In a preliminary step, the short exposures recorded by
CP20 were cleaned for photon-centroiding electronic artifacts and then co-added to generate an equivalent  long
exposure of 99s for P Cyg and 370s for 59 Cyg. Despite the Coud\'e configuration of
the 1.52m telescope, rotation of the field between the two sequences was found negligible. We
checked for non-linear effects introduced by the so-called photon-centroiding hole and camera
saturation (\cite{thiebaut94}). It appears that the incoming flux is much below the saturation limit,
the double-photon occurrence is absent. The total number of detected photons for the two
images are very similar: 481563 for P Cyg and 416309 for 59 Cyg. A coherent peak and a broken ring, featuring a triple coma aberration, are clearly
visible, but 59 Cyg's image is sharper, indicating that P Cyg's
envelope is possibly resolved. This is quite visible in Fourier space,
as shown in Fig. \ref{mod}. The envelope of P Cyg clearly appears as the well
  resolved low frequency part in the visibility curve. High angular
  information is obtained up to about 80 \% of the cut-off
  frequency. The visibility curve is indeed clearly dominated by noise
  beyond that limit.\\
\begin{table}[t] \caption[]{Turbulence characteristics derived from
    the wavefront sensor data of the adaptive optics system. Atmospheric parameters are given for the visible band.} 
\begin{center} 
\begin{tabular}{|l|c|c|} \hline 
Star name & P Cyg & 59 Cyg \\ \hline 
Time & 20:30 TU & 21:23 TU \\ \hline 
Average intensity & 51.567 & 42.830 \\ in sub-apertures (photons) & & \\ \hline 
$r_0$ (cm), open loop & 5.83 & 5.92 \\ calculated with $L_0$ & & \\ \hline
$r_0$ (cm), open loop calculated & $\sim$5.4 & $\sim$5.4 \\ with $L_0$ and Zernike coef. & &\\ \hline 
$L_0$ (m) & 2.96 & 2.57 \\ \hline 
Zernike coef. (2,3), $\sigma^2$ ($rad^2$) & 70 & 40 \\ \hline 
Zernike coef. (4,44), $\sigma^2$ ($rad^2$) & 10 & 9  \\ \hline 
\end{tabular} 
\end{center}
\end{table}
Since 59 Cyg has been observed 53 minutes later than P Cyg, a variation of atmospheric
conditions and/or a poor correction could be responsible of a noticeable change in the Point Spread
Function (PSF) shape. From open/close loop data recorded by the wavefront sensor of BOA, seeing
conditions have been derived by one of us (F.C.) : intensity in sub-apertures, Fried parameter
evolution, noise evolution, Zernike coefficient variance evolution, etc... Some of these relevant
parameters are summarized in Table 1 for both stars. In particular the Fried parameter ($r_0$) exhibits
no temporal variation and remains stable around 5.4cm between the 2 data sequences. Also, high
order correction (Zernike coef. (4,44)) have similar variances in both cases, indicating similar
adaptive mirror corrections for P Cyg and 59 Cyg. The largest
difference appears for the tip-tilt corrections (Zernike coef. (2,3)
of the Table 1). However we confirm that during the observations no saturation was detected, and we expect these
 modes be correctly compensated for both stars. Furthermore,
parameters such as turbulent layers altitude, mean isoplanetic angle
and wind velocity indicate a very good stability of the atmospheric
conditions during the two records.\\ 
 Moreover, the photon flux stability can be
checked on the short exposures statistics. For 59 Cyg, the flux is stable with a perfect Poisson
variance. P Cyg's behavior is roughly the same, except some localized empty exposures, due to
electronic saturation, which have been removed. \\  This preliminary analysis permits to assume that 59 Cyg can be confidently considered as P Cyg's PSF and a successful deconvolution becomes possible.\\

\section{Image reconstruction}
\subsection{Deconvolution of the raw image of P Cyg}
A deconvolution of the image of P Cyg using 59 Cyg as the PSF has been made 
using basically the iterative Richardson-Lucy algorithm (RL)
(\cite{richardson72}, \cite{lucy74}) to which some improvements have
been applied as we shall describe thereafter. 
To improve the result, a pre-processing of the image was first
performed. All images were smoothed using a moving average
window of 4$\times$4 pixels. 
To get rid of the effects of the coronagraphic mask (clearly visible
in Fig. \ref{figun} at the right of P Cyg's image), only a central 
region of 0.9$\times$0.9 arcsec of the smoothed images was conserved. 
To make the deconvolution procedure faster, the number of points of
the images was further reduced by a factor 16. The resulting images of 32$\times$32 pixels remain 
correctly sampled (0.028 arcsec
compared with the theoretical Shannon limit of 0.044 arcsec for the
1.52m telescope operated at 656.3nm).\\
The RL algorithm was then applied to these pre-processed images using the well-known iterative procedure:
\begin{equation}
{x}^{k+1}(r) = {x}^{k}(r) \cdot  h(-r) \otimes \frac{ y(r)}{h(r) \otimes {x}^{k}(r)}
\end{equation}
where $\otimes$ is the convolution symbol, $r$ denotes the two-dimensional spatial position, $y(r)$ denotes the image of P Cyg, 
$h(r)$ the image of 59 Cyg taken as the PSF, and where ${x}^{k}(r) ,  
{x}^{k+1}(r)$ are the  reconstructed images at the iteration $k$ and
$k+1$.\\
\begin{figure}[t]
\centerline{\resizebox{5.0cm}{!}{\includegraphics{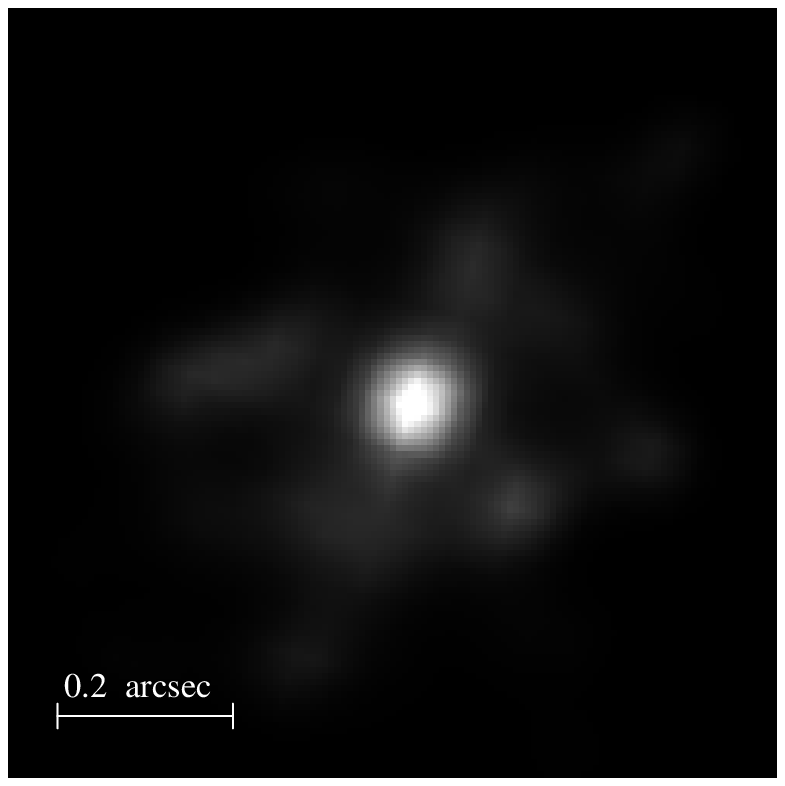}}}
\centerline{\resizebox{5.0cm}{!}{\includegraphics{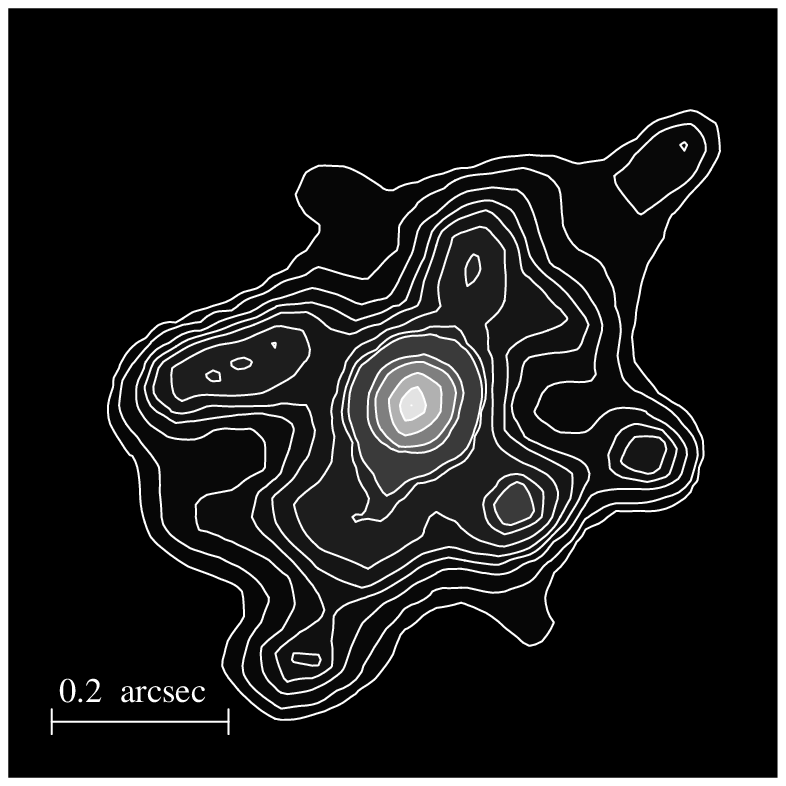}}}
\vspace{0.2cm} \caption[]{Reconstructed image of P Cyg with the RL algorithm stopped at iteration 245 . Top: Representation in gray
levels in a linear scale. Bottom: Representation in contour
plot. Contours levels are not equally spaced and correspond to : 100,
80, 60, 40, 30, 15, 12.5, 9, 6.5, 5, 3.5, 2, 1.2 percent of the
image's maximum. North is at the top and East is at the right of the
images. This corresponds to a rotation of 104.7$^{\circ}$ clockwise
with regards to Fig. \ref{figun}.} \label{figtrois} \end{figure}  
To limit the instability that appears in the solution, due to 
the amplification of noise, we stop the iteration number by using a 
comparison between the Fourier Transform of the reconstructed object 
at the iteration $k$ and the Fourier transform of the image 
reconstructed by a Wiener filter  (\cite{lanteri98}, \cite{lanteri299}). 
The main difficulty is then shifted to a correct determination of the Wiener filter. 
The comparison lead to choose an iteration number of 245. 
The result of the deconvolution is given both in gray levels and in
contour plot in Fig. \ref{figtrois}.
To make the envelope clearly visible, the contour levels are not equally
spaced. 
The general pattern is that of a bright star, not resolved by the 152 cm telescope, 
surrounded by an extended envelope with bright spots. \\
These same
results were obtained when we used for the reconstruction the RL
algorithm regularized by a Tikhonov term, and more specially using the
Laplacian operator (\cite{lanteri99}). The result of this deconvolution is presented in
Fig. \ref{figquatre}  for the iteration 250 with a regularization factor equal to 0.01.
 \begin{figure}[t]
\centerline{\resizebox{5.0cm}{!}{\includegraphics{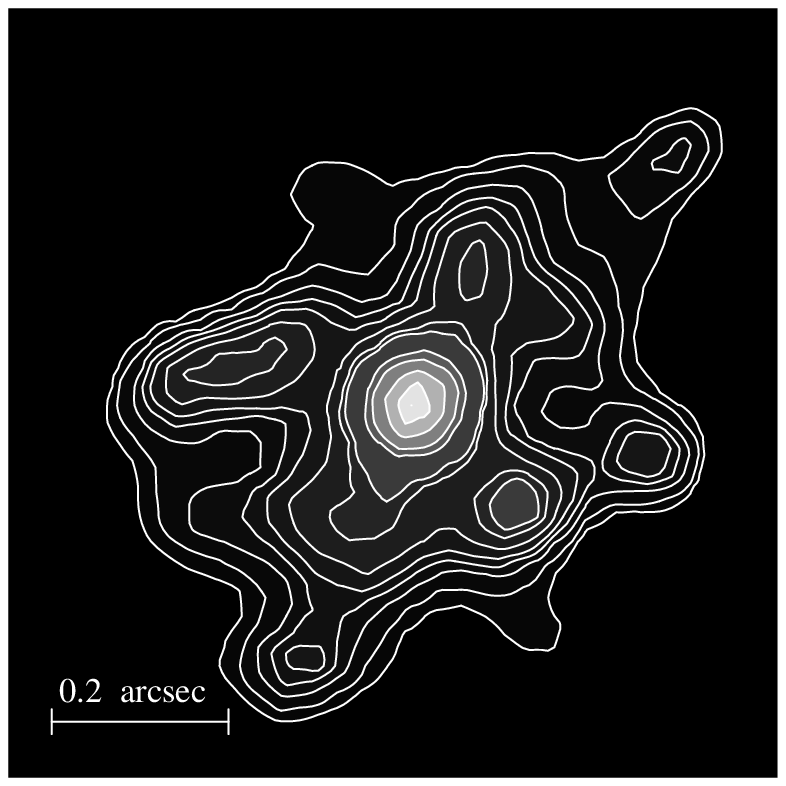}}}
\vspace{0.2cm} \caption[]{Reconstructed image of P Cyg with the RL
  algorithm regularized by the Laplacian operator. Contours levels are
  not equally spaced and correspond to: 100, 80, 60, 40, 30, 15, 12.5, 9, 6.5, 5, 3.5, 2, 1.2 percent of the image's maximum. North is at
the top and East is at the right of the image. } \label{figquatre}
\end{figure}
\begin{figure}[t]
\centerline{\resizebox{7.0cm}{!}{\includegraphics{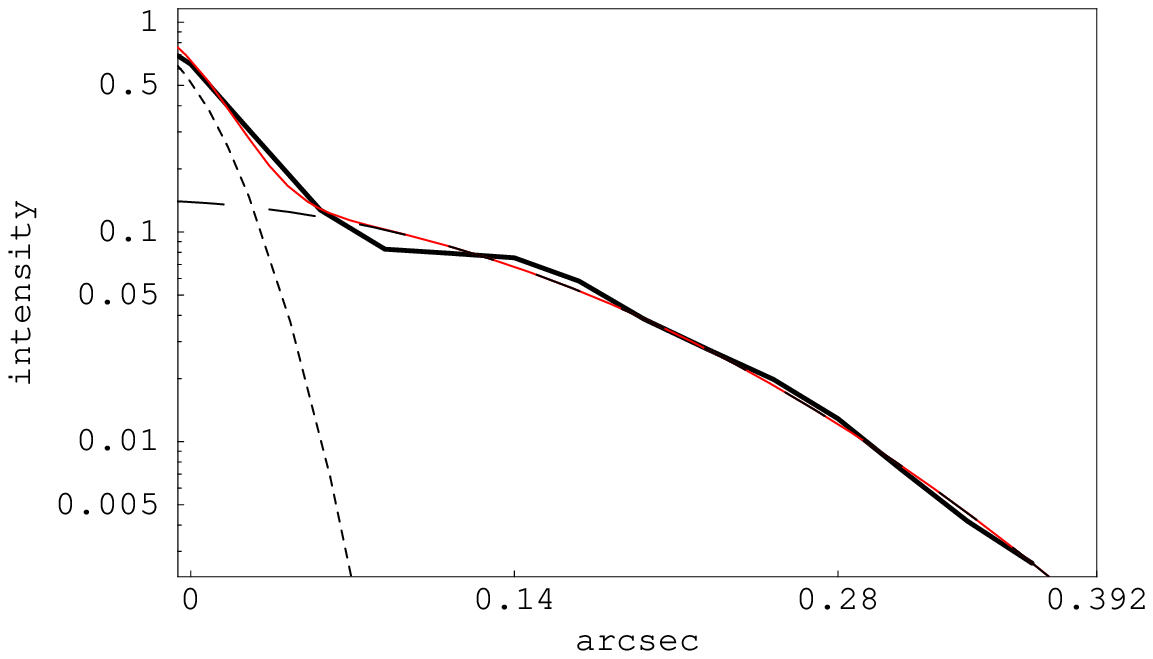}}}
   \caption[]{Thick line: Radial profile of the object reconstructed by RL stopped at iteration 245 (image normalized to one). Dashed lines: Gaussian curves. Thin line: Sum of the two Gaussian curves. Intensity is in logarithmic scale.} \label{figcinq} \end{figure} 
Neglecting the fine structure of the envelope, a normalized circular 
average of the image centered on the star may be fitted by the sum 
of two Gaussian curves, of the form: $a {e}^{-b |r|^{2}} + (1-a) {e}^{-b' |r|^{2}}$ 
where a $\simeq$ 0.85, b $\simeq$ 618.2 and b' $\simeq$ 25.9 if we
express r in arcsec. This corresponds to an equivalent width of
the envelope of about 0.4 arcsec ($\frac{2}{\sqrt{b'}}$).
The radial curve is shown in a semi-logarithmic scale in
Fig. \ref{figcinq}.
In this model the total integrated flux produced by the envelope (r = (1-a)b/ab') 
was found to be about 4 times larger than the one produced by the central
star. This ratio is not an absolute parameter; it depends of course of
the spectral bandwidth of the experiment (Fig. \ref{figdeux}).\\
\begin{figure}[t]
\centerline{\resizebox{8.0cm}{!}{\includegraphics{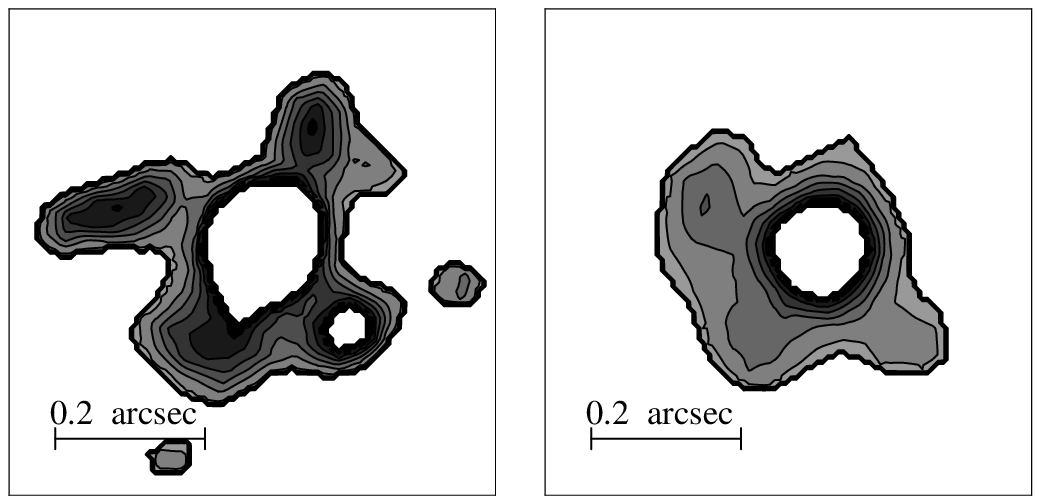}}}
   \caption[]{Representation at the same scale of the reconstructed image of the Fig. \ref{figquatre} (left) and of the PSF (right) in contour plot.
  The white parts are due to the threshold and the saturation applied
  to both images: intensities lower than 6 \% and higher than 15 \%
  of the maximum are made white. North is at the top and East is at
  the right of the images.}
\label{figcinqavant} \end{figure}
 We may conclude from this first analysis that the envelope of P
 Cyg is well resolved by the central core of the PSF of the 1.52 m
 telescope, corrected by the adaptive optics system. However, the
 envelope remains comparable in size with the residual halo of the
 PSF. This assumes at least that we are free of  anisoplanatism
 problems (\cite{fusco00}); this absence of variation allows the use
 of conventional deconvolution methods.\\The question then may arise
 whether some of the fine structures we discovered in this envelope
 may be due to artifacts of the reconstruction process. Among others
 problems, a very important one is a possible variation of the PSF
 during the experiment (from P Cyg to 59 Cyg), as discussed recently
 by Harder and Chelli (\cite{chelli00}). These authors show that a
 local non-stationary turbulence may produce strong residual
 aberrations (clearly visible in the first diffraction ring of their figure 17). At worst, can we
imagine that the observed structures in our image result only from variations in the PSF, from P Cyg to 59 Cyg?\\
A comparison between the reconstructed image and the PSF is made
 in Fig. \ref{figcinqavant}. To make the structures more visible, we used a representation similar to that of Harder and Chelli
 (\cite{chelli00}). The images are negatives of intensities and the representation uses threshold and saturation. Doing so, a strong
(white) secondary maximum appears in the image of P Cyg, and two lower
 ones remain far away from the core. Moreover, Fig. \ref{figcinqavant}
 clearly evidences that the central (white) surface of the reconstructed object is larger than that of the PSF (this will be interpreted
in the next section as an effect of the envelope). It seems difficult
to explain that such a structured image is only the result of
 variations of the PSF. However, it is also difficult to ascertain that
 our reconstructed image is free of any residual error (much more data
 would have been necessary for that). To strengthen our confidence in
 the fine details of the reconstructed image of the envelope, we have
 implemented a series of processing and tests. They are described in
 next sections 3.2 and 3.3, and in the Appendix.
\begin{figure}[t]
\centerline{\resizebox{5.2cm}{!}{\includegraphics{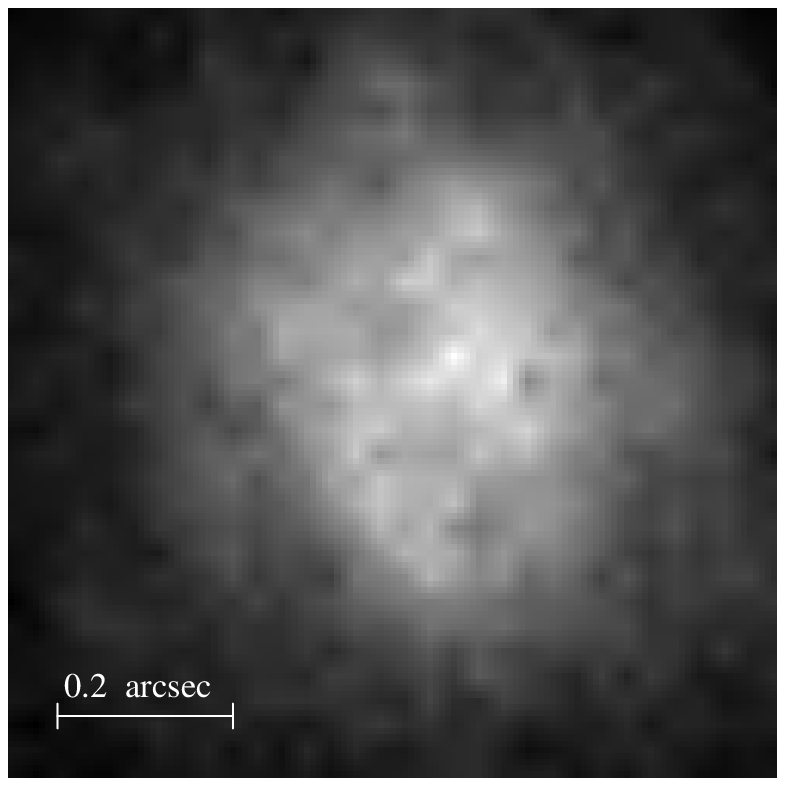}}}
\centerline{\resizebox{5.2cm}{!}{\includegraphics{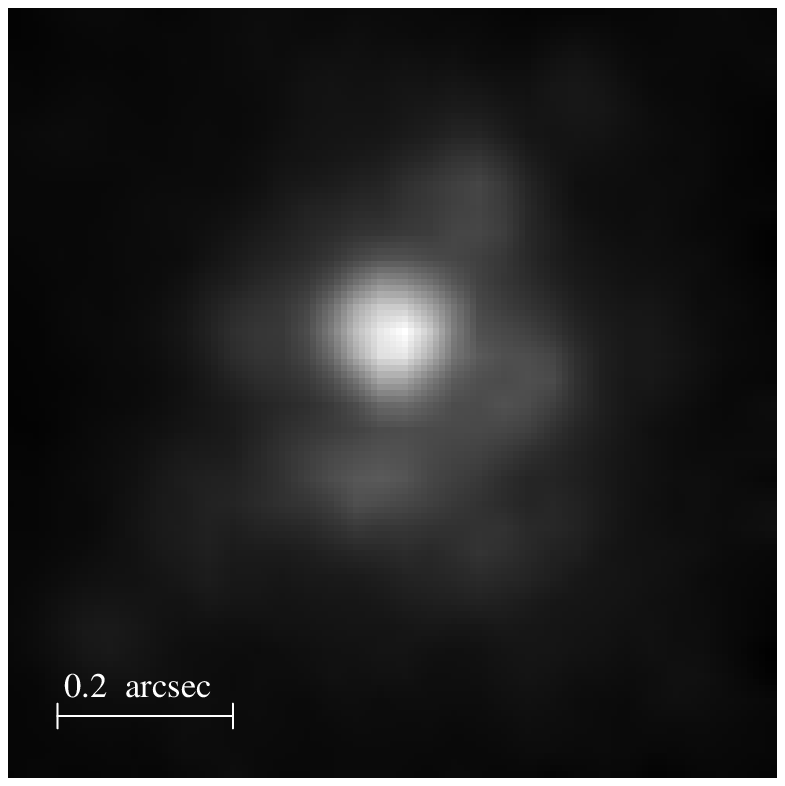}}}
\vspace{0.2cm}
  \caption[]{Representations in gray levels of the residual blurred envelope $e(r) \otimes h(r)$ (top) and of the fraction $\gamma h(r)$ of the image of 59 Cyg subtracted from P Cyg (bottom). The
   sum of these two images gives back the observed image of P Cyg. The
   orientation is that of Fig. \ref{figun}.} \label{figsix} \end{figure}
\subsection{Deconvolution of the envelope - alone of P Cyg}
 We have further processed the data assuming that P Cyg observed
in  $H_{\alpha}$ may be fairly represented by an unresolved star
 surrounded by an extended envelope. Let us write P Cyg as
 $e(r) + \gamma\delta(r)$, where $e(r)$ is the envelope and $\gamma$
 the relative intensity of the star defined by the Dirac function
 $\delta(r)$. The observed image is then modeled as $y(r) = e(r)
 \otimes h(r) + \gamma h(r)$.
 We have implemented a somewhat heuristic procedure that consists of
 subtracting $\gamma h(r)$ from $y(r)$ to obtain $e(r) \otimes
 h(r)$. The procedure has some similarity with what is made in the
 algorithm CLEAN (\cite{hog74}), and the algorithm of Lucy (\cite{lucy94}). In practice, it consists of
 subtracting an appropriate fraction of the image of 59 Cyg correctly
 shifted from P Cyg. The parameters of this subtraction are determined
 to leave a smooth pattern for $e(r) \otimes h(r)$, with no remaining
 bump or hole due to the unresolved star. We found that a fraction
 $\gamma$ = 20 \% of the PSF ($h(r)$), shifted of a displacement (X =
 - 0.3 pixels, Y = + 1.3 pixels) was to be subtracted from P Cyg
 ($y(r)$). This was done using an interpolation of the images with Mathematica
 (\cite{wolfram95}). Fig. \ref{figsix} shows the residual
 blurred envelope $e(r) \otimes h(r)$ (top) and the fraction of the PSF
 subtracted to the image of P Cyg $\gamma h(r)$ (bottom).
This procedure, performed independently of the above deconvolution and
 parametric estimation of the star plus envelope, lead also to the
 same ratio 4 for the energy of the envelope relative to the central
 star.\\
\begin{figure}[h]
\centerline{\resizebox{5.2cm}{!}{\includegraphics{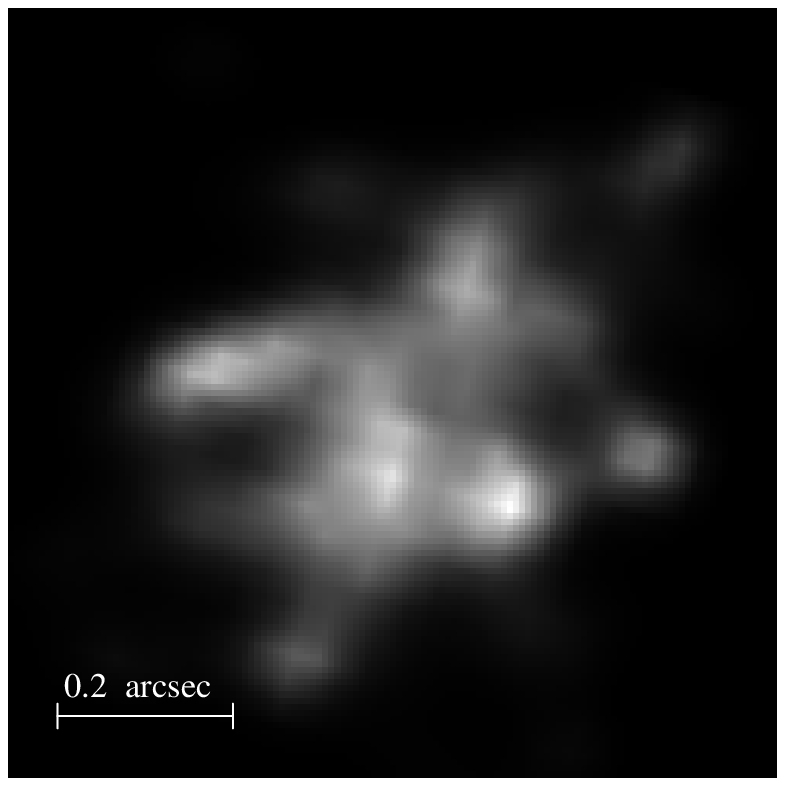}}}
\centerline{\resizebox{5.2cm}{!}{\includegraphics{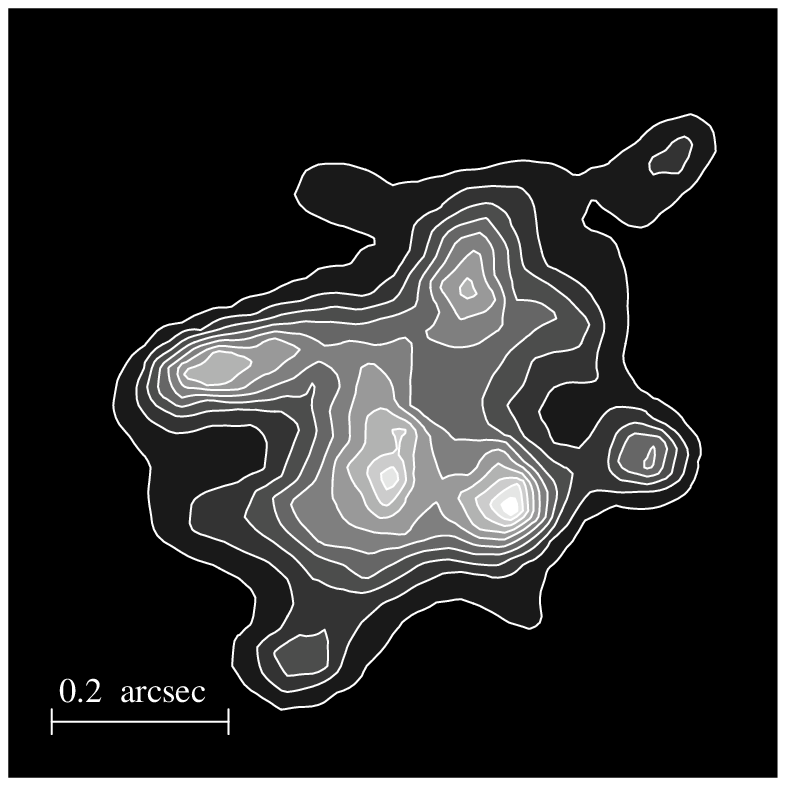}}}
\vspace{0.2cm} \caption[]{Image reconstruction of the envelope of P Cyg in gray levels (top) and in contour plot (bottom). Contours levels are equally spaced. North is at the top and East is at the right of the images.} \label{figsept} \end{figure}\\
We then processed the residual blurred envelope $e(r) \otimes h(r)$
using the same RL algorithm stopped at the same iteration number
245. The result is given both in gray levels and in contour plot with
linear contour spacing in Fig. \ref{figsept}.  
The image of the envelope is fully consistent with what was obtained in the raw deconvolution of Fig. \ref{figtrois}. 
The bright spots are all found at the same position. Moreover, there
is a bright spot very close to the star clearly visible in this
representation in the South-West direction; it was only perceptible as a small deformation of the central star in Fig. \ref{figtrois}.\\
\subsection{Analysis of the quality of the deconvolution}
\begin{figure}[t]
\centerline{\resizebox{5.2cm}{!}{\includegraphics{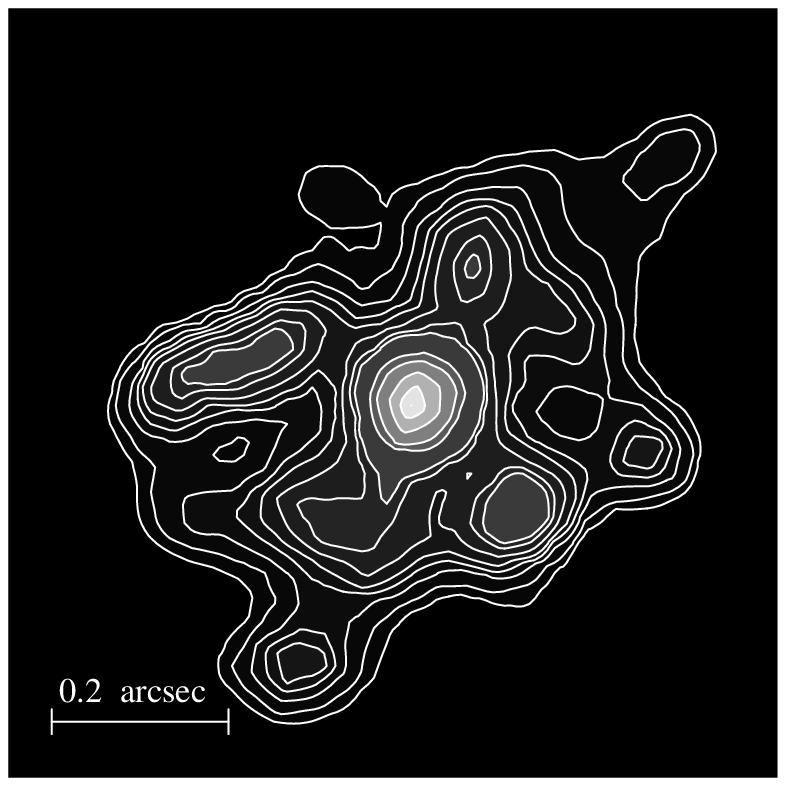}}}
\vspace{0.2cm} \caption[]{Reconstructed image of P Cyg with the ISRA algorithm stopped at iteration 303. Contours levels are not equally spaced and correspond to: 100, 80, 60, 40, 30, 15, 12.5, 9, 6.5, 5, 3.5, 2, 1.2 percent of the image's maximum. North is at the top and East is at the right of the image. } \label{fighuit} \end{figure}

 We have implemented a few more computations to check the quality of
 the results we give in this paper. The whole deconvolution procedure was also performed using the Image
 Space Reconstruction Algorithm (ISRA) (\cite{daube86}) instead of the
 RL algorithm. The comparison with the image reconstructed by the
 Wiener filter leads us to the iteration 303. The reconstructed image is presented in Fig. \ref{fighuit}.
 The two algorithms gave almost the same results which
 reinforced us in the confidence we have in the deconvolution results
 given in this paper. \\
At that point we may conclude that the deconvolution was
carried out taking into account the problem of noise; we also made use of
an a
 priori model, by assuming that P Cyg was the sum of an
 unresolved star and an envelope. A significant question remains: up
 to what precision can
 we trust 59 Cyg as an accurate PSF for P Cyg? Several elements
 enable us to give a positive answer to that question: the seeing
 conditions were similar for both observations, and we found the same
 ratio of the flux envelope/star (about 4) before and after
 deconvolution. It would have been very convincing to have a series of
 observations of P Cyg and a reference, eventually with different
 seeing conditions, and have all the results that converge towards a unique
 solution. In fact, we made an elementary test that
 consisted in dividing the long exposure of P Cyg in two
 sequences. The same operation was applied to 59 Cyg. We have then made
 a deconvolution of these four resulting images. The results are very similar
 to those obtained with the deconvolution presented in Fig.
 \ref{figtrois} and are not reproduced here. In an alternative way, we give in the appendix the results of
 a series of additional tests that tend to confirm 59 Cyg as a correct
 PSF.\\
 At that point, we think that we obtained all that allowed the present data. Of course, futures observations,
 with possibly a larger telescope, will be very usefull to confirm
 our first results and to further precise the
  morphology of this object.

\section{Discussion} 
The present work follows a long series of observations to track the history
of P Cyg's giant eruptions in 1600 and 1655. Following the first attempts
in this direction (\cite{feibelman95}), it clearly appeared that the star
to its nebula brightness ratio constitutes a challenging 
obstacle
for the optical mapping of the nebular morphology. Indeed only few
observations, with heterogeneous angular resolutions are reported in the
literature. Leitherer and Zickgraf (\cite{leith87}) first published the
detection of P Cyg's extended nebulosity using CCD imaging. Later,
Johnson et al. (\cite{johnson92}) reported the detection of forbidden emission lines due to nitrogen 
enriched material at 9 arcsec.
The first coronagraphic imaging of P Cyg from the ground
was obtained by Barlow et al. (\cite{barlow94}) from high resolution long-slit data.
Barlow and co-workers discovered the presence of a 22 arcsec nearly
circular shell which presents bright condensations of about 2 arcsec wide, mostly in the 
North (their plate scale is of 0.255 arcsec per pixel where a 4.3 arcsec 
occulting
strip was used). STScI coronagraph imaging, using a 4.8 arcsec occulting disk, shows almost 
the same features (\cite{nota95}).\
From another point of view P Cyg constitutes an ideal 
target for testing new and different imaging techniques. Among these
there are the spectral-line image sharpening
techniques SCASIS (\cite{devos94}), the AMOS adaptive optics observations
(\cite{morossi96}),
and occulting mask imaging (4 arcsec) with the new MOMI instrument for wide
field imaging
(\cite{oconnor98}). In the latter case the authors suggest highly 
asymmetrical features at 3-4pc from the star (7') probably associated 
to previous mass-loss events (\cite{meaburn99}).
At the same time, radio imaging now offers approximately the same panel of 
field and
spatial resolution as optical imaging making the comparison of optical and 
radio maps possible. Indeed, sounding different scales in radio wavelength 
can easily be achieved by
changing the baseline configuration of radio arrays (\cite{skinner98}). 
Using
this possibility
Skinner and co-workers have compared radio maps to Barlow's coronagraphic 
image
of P Cyg
(\cite{barlow94}). These authors  claim that the emissive regions in radio 
and
visible
are roughly the same, although this comparison is further complicated by the
coronagraphic
mask and the telescope diffraction pattern. They conclude that both radio 
and
optical
maps exhibit the same details having the same physical origin: i.e. dense 
clumps
overtaken
by the faster wind and heated by shocks.\\
Low and intermediate spatial resolution images suggest a global spherical
expanding envelope
but clumpiness is present in each case. This trend is also
present at higher spatial resolution of the 250km MERLIN centimetric network
(\cite{skinner97})
approaching the submilliarcsecond observations of the GI2T optical
interferometric (\cite{vakili97}).\\
In this context our present reconstructed image of P Cyg's environment in
H$_{\alpha}$ presents,
as expected, strongly clumped features within the 1 arcsec field of view
(0.011pc with
D=1.8Kpc). More than 6 enhanced emission clumps are counted with our 0.05
arcsec spatial resolution in a nearly 0.6 arcsec region superimposed
to the unresolved central star. The mean size 
of the clumps is roughly
0.08 arcsec which is the angular limitation of the 152cm OHP telescope. These
results agree well both
in size and morphology with MERLIN observations, at nearly the same
resolution.
The typical diameter of emitting regions for MERLIN is 0.4 arcsec (0.13 
arcsec
for the core),
and amazingly comparable to the optical structures (Fig. \ref{figsept}). In
this same figure, a North-East/South-West
preferential axis
appears in the H$_{\alpha}$ image due to the grouping of 
the
clumps, the distribution being otherwise rather uniform. The same
orientation was also pointed out in 
SCASIS
observations
at a lower spatial resolution (\cite{devos94}).
Note that in our reconstructed image a bright feature is located at 80 mas South-East of 
the
central star.
We can speculate on its relation to the local strong emission discovered at 
0.8 mas from the
star in august 1994 by the GI2T interferometer (\cite{vakili97}) although the E-W absolute 
position
of the latter emission was not given by this instrument. If this
scenario holds, this position some 3.2 
years after implies
a projected velocity around 110km/s. Taking into account the radial velocity 
and uncertainties 208$\pm78$km/s obtained by Vakili et al. (\cite{vakili97}), this
projected velocity is to be expected for a clump nearly on the line of sight 
ejected three years earlier with the terminal velocity, and thus, compatible
with the GI2T observations. Although the possible physical relation
of 1994 blob and 1997 clump remains to be robustly settled,
interferometric and AO imaging repeated in the future, should enlight
such scenario.\\
At the present, only radio observations by Skinner et al. (\cite{skinner97}) present confident temporal
variations.
In the two 6-cm MERLIN images taken in a 40 days interval, impressive 
changes
were observed,
corroborated by VLA observations. In the observed region, the wind velocity
suggests a 2 years
dynamical time scale, which can hardly be compared to the 6-cm flux 
variations.
On the other hand,
the recombination time scale for hydrogen atoms ($1.2 10^5/ n_e$ in years) 
is
shorter, but
not sufficient, $\sim$160 days considering a characteristic $n_e$ of $2.8 
10^5
cm^{-3}$ at
0.07 arcsec. This short time scale puts strong constraints on the electron
density, which has
to be four times larger than the surrounding envelope material. The 
clumpiness
can explain such a
time evolution if the structures are sufficiently small and dense, or if the
shocks between the
wind and the clumps are strong enough.  The question is whether increasing
resolution would reveal
the same clumpiness, and if activity observed in optical and radio 
wavelengths
are closely
correlated.\\
Some questions arise. How can these small
scale clumps be
related to the 2 arcsec ones observed in the Barlow's images at 3 arcsec 
from
the star? How
can they survive over such a long distance? Do they reappear at the
location we detect them? \\
A challenging issue is now understanding the connection between the different spatially resolved
structures and their
scales which needs the monitoring of the clumpiness from the star to the interstellar 
medium,
and constrains
the mass ejection dynamics. Therefore, our present observations are the 
first
attempt to prove
that an optical monitoring of the clumpiness in inner region of P Cyg's mass
loss is observationally
feasible. As previously pointed out (\cite{vakili97}), the intermediate 
regions
of P Cyg's wind,
from a few stellar radii to a few parsecs can be sounded by means of AO plus
coronagraphic
imaging from the ground, in relation with radio observations. The dynamical 
time
scale for
optical interferometry is around the month, but a temporal monitoring of P 
Cyg
by this technique
requires both higher sensitivity and larger numbers of baseline orientations  
due
to the complex structures which occur at different scales. For larger distances, the 
recombination
time scale in the
clumpy wind should produce large effects as detected in radio, and AO 
becomes
the perfect technique
to follow such activities. The brightness of P Cyg and its evolutionary 
time
scale allow the
development of a multi-site and multi-wavelength observations campaign, 
using
AO, optical and
radio interferometry to get a unified picture of P Cyg's environment 
physics.
\appendix
\section{Tests on the PSF}
We have made a series of tests that tends to strengthen our confidence
in the 59 Cyg's image as a correct PSF for P Cyg. These tests are
negative tests in that sense that if the PSF fails to succeed one, it
should be rejected as a bad PSF.\\
Let us assume that 59 Cyg is a good unbiased representation of the
true PSF $h(r)$. Differences between the blurred reconstructed image
$b(r) = h(r) \otimes x(r)$ and the observed data $y(r)$ must then be
dominated by statistical noise fluctuations, with no bias term. In this relation
$x(r)$ is our best reconstructed image, as shown in Fig. \ref{figtrois}, and
$h(r)$ the PSF 59 Cyg of Fig. \ref{figun} of the body of the paper. We assume
that the noise comes from a photo-detection process, and that $y(r)$ is
a realization of the Poisson process of mean $b(r)$.
\subsection{Test 1: Poisson-Mandel transform}
The first test we have performed is a basic one, not very sensitive to
the exact value of $h(r)$, but that must be verified in any case. Let us
denote $y$ and $b$ the values taken by $y(r)$ and $b(r)$. The total
probability theorem (\cite{papou84}) allows us to write the unconditional
probability $P(y)$ of $y(r)$ as the sum of $P(y/b)\times P(b)$ for all $b$
values. For the Poisson process, the conditional probability of $y$
assuming $b$ is $P(y/b) = \frac{b^y}{y!} \exp{-b}$ where $y$ is an
integer (the number of photons) and $b$ a continuous value.\\
\begin{figure}[t]
\centerline{\resizebox{7cm}{!}{\includegraphics{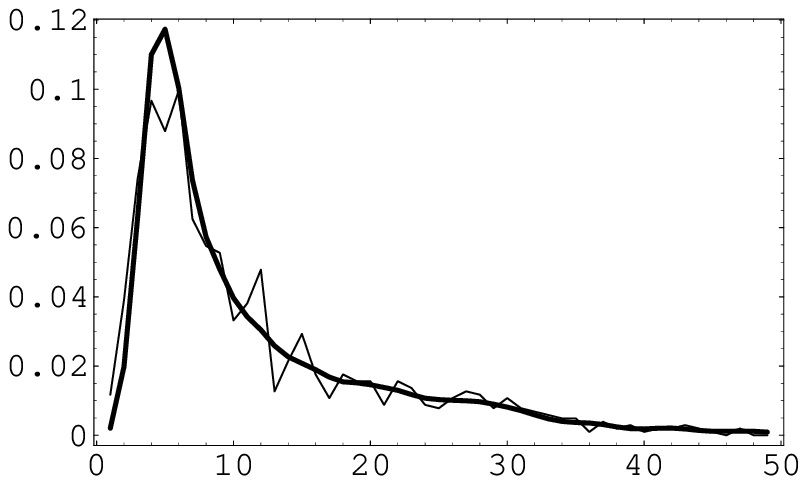}}}
\caption[]{Thick
    line:Poisson-Mandel transform $P(y)$ of $P(b)$ calculated using relation \ref{pois}. Thin line: Histogram of the values of $y(r)$. A good agreement between the two curves is observed.} \label{histo}\end{figure}
As a
consequence, $P(y)$ is the Poisson-Mandel transform (\cite{mandel59}, \cite{mehta70}) of $P(b)$:
\begin{equation}
 P(y) = \int_{0}^{\infty} P(b) \frac{b^y}{y!} \exp{-b} \hspace{0.2cm}db \label{pois} 
\end{equation}
We have verified that our data correctly obeys relation \ref{pois}. We
have taken for $P(b)$ the histogram of $b(r)$, applied the
above transformation and compared the result
$P(y)$ with the direct histogram of the values of $y(r)$. The
comparison is shown in Fig. \ref{histo}. The results are consistent with the
data.
\begin{figure}[ht]
\centerline{\resizebox{5.2cm}{!}{\includegraphics{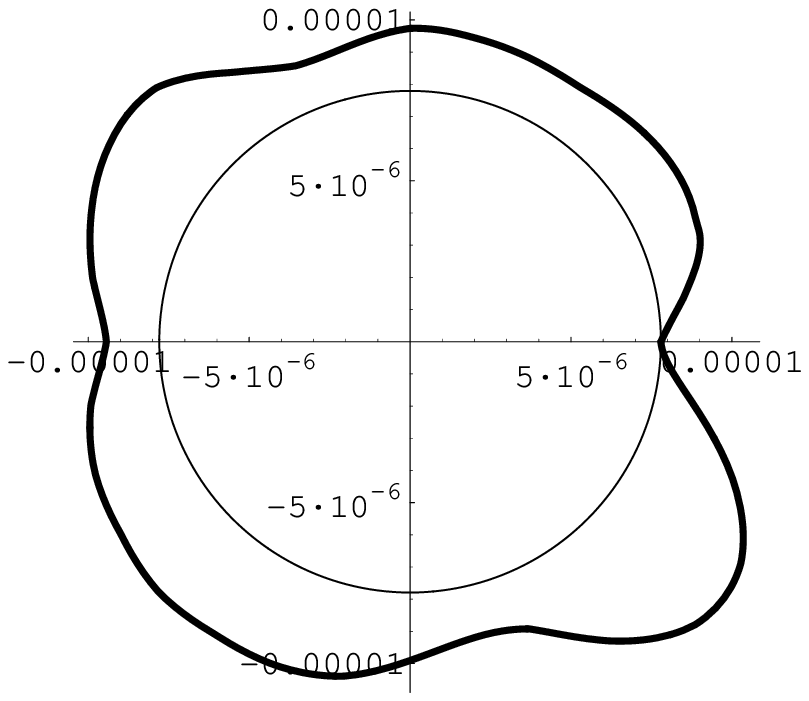}}}
   \caption[]{Thick
   line: Polar plot of the Euclidean distance between $y(r)$ and
 $b_{\theta}(r)$ for $\theta$ varying between 0 and 2$\pi$. Thin
 line: Circle with a radius of the minimum of the Euclidean distance
  between  $y(r)$ and $b_{\theta}(r)$. The minimum of the Euclidean
  distance is obtained for $\theta$ = 0.} \label{polar}\end{figure}
\subsection{Test 2: Rotations of the true PSF} 
The second test we have performed was to check the correctness of the
small departures from circular symmetry of 59 Cyg. We use as the PSF
$h_{\theta}(r)$ an image of 59 Cyg rotated of an angle $\theta$. Then the deconvolution procedure is carried out as previously
and leads to an image $x_{\theta}(r)$.
The blurred image $b_{\theta}(r)$
is obtained as the convolution of $x_{\theta}(r)$ and
$h_{\theta}(r)$, and we finally compute the Euclidean distance between
$y(r)$ and $b_{\theta}(r)$. The results are shown in Fig. \ref{polar}
in a polar plot for $\theta$ varying from 0 to 2 $\pi$. 
The original PSF gives the
best result. If the deviation from circular symmetry were purely
random, the goodness of the deconvolution would not be affected by
this rotation. Secondary minima of the curve appear for apparent
symmetries of the PSF.\\
\begin{figure}[h]
\centerline{\resizebox{4.3cm}{!}{\includegraphics{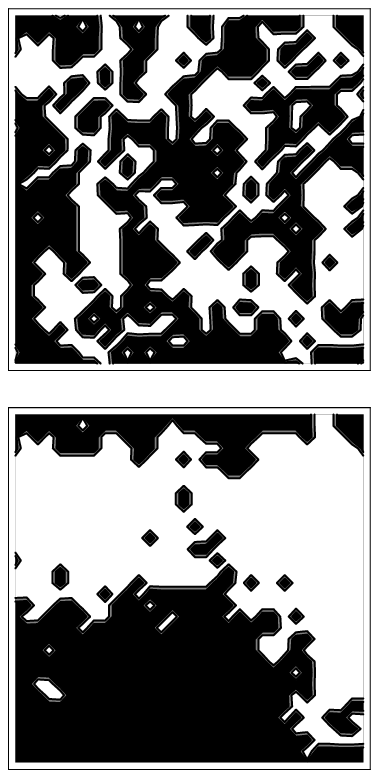}}}
  \caption[]{Representation of the sign of the difference $y(r)- b_{\theta}(r)$ for no rotation of the PSF (top) and for a rotation of $\pi/2$ (bottom).  This later image is not a particular case, and is characteristic of what can be obtained when the PSF is rotated.} \label{signe}\end{figure} \\
Representations of the difference between $y(r)$ and
$b_{\theta}(r)$ are shown in Fig. \ref{signe}. For the sake of clarity we have represented the
sign of the difference $y(r) - b_{\theta}(r)$. 
For $\theta = 0$, we get a
speckle-like pattern, roughly uniform over the whole image. For other
values of $\theta$, this difference shows large patterns that indicates
regions over which $b_{\theta}(r)$ does not correctly matches
$y(r)$. These trends clearly show that $ h_{\theta}(r)$ for $\theta \neq 0$
is a biased version of $h(r)$.\\
As a conclusion, 59 Cyg passed the tests and could not be rejected
as a bad PSF.
\begin{acknowledgements}
We wish to acknowledge A. Labeyrie for having encouraged the present
work. P Cyg's observations were done using the BOA Adaptive Optics
provided by ONERA to us. The manuscript benefited from discussions and
critics of J.P. V\'eran, J. De Freitas Pacheco, G. Ricort and the GI2T team.
\end{acknowledgements}

 \end{document}